\documentclass[twocolumn,showpacs,preprintnumbers,prl,amsmath,amssymb]{revtex4}

\usepackage{graphicx}
\usepackage{dcolumn}
\usepackage{bm}
\usepackage{color}
\usepackage{mathrsfs}

\newcommand{\llgg}{{\tiny \raisebox{0.0ex}{$<$}\hspace{-0.75em}\raisebox{1.5ex}{$>$}}}

\begin{document}

\preprint{}

\title{Nonequilibrium Green's Function Approach to Phonon Transport in Defective Carbon Nanotubes}
\author{Takahiro Yamamoto}
\author{Kazuyuki Watanabe}
\affiliation{Department of Physics, Tokyo University of Science, 1-3 Kagurazaka, Shinjuku-ku, Tokyo 162-8601, Japan}
\affiliation{CREST, Japan Science and Technology Agency, 4-1-8 Honcho Kawaguchi, Saitama 332-0012, Japan.}

\date{
\today
}
             
\begin{abstract} 
We have developed a new theoretical formalism for phonon transport in nanostructures using the nonequilibrium phonon Green's function technique and have applied it to thermal conduction in defective carbon nanotubes. The universal quantization of low-temperature thermal conductance in carbon nanotubes can be observed even in the presence of local structural defects such as vacancies and Stone-Wales defects, since the long wavelength acoustic phonons are not scattered by local defects. At room temperature, however, thermal conductance is critically affected by defect scattering since incident phonons are scattered by localized phonons around the defects. We find a remarkable change from quantum to classical features for the thermal transport through defective CNTs with increasing temperature.
\end{abstract}

\pacs{44.10.+i, 61.46.Fg, 63.20.Mt}
 
\maketitle
The discovery of the quantization of phonon thermal conductance as well as quantization of electrical conductance has had a great impact on mesoscopic and nanoscopic physics~\cite{rf:rego,rf:ange,rf:blen}. Since the first observation of quantized thermal conductance $\kappa_0\equiv\pi^2k_B^2T/3h$ in dielectric mesoscopic wires by Schwab {\it et al.}~\cite{rf:schw}, significant effort has been expended on exploring new materials exhibiting $\kappa_0$. Carbon nanotubes (CNTs) are expected to be potential candidates to measure quantized thermal conductance~\cite{rf:yama1,rf:mingo} because they are ideal one-dimensional phonon conductors with a large phonon-mean-free path~\cite{rf:mingo,rf:hone,rf:yu}. Recent sophisticated experiments have in fact succeeded in measuring quantized thermal conductance in CNTs~\cite{rf:chui}. \par

Recently, concern has been raised that various intrinsic features of pure CNTs are lost because of the presence of defects in synthesized CNTs. According to recent molecular-dynamics (MD) simulations, the thermal conductivity in CNTs at room temperature decreases dramatically with increasing defect density~\cite{rf:che,rf:maruyama,rf:zhang,rf:kondo}. Similar to room-temperature thermal transport in defective CNTs, low-temperature thermal transport in CNTs is predicted to be markedly affected by defects, and quantized thermal conductance might be destroyed due to the presence of the defects. To the best of our knowledge, the influence of defect scattering on low-temperature thermal transport through defective CNTs has not been studied thus far. The central aim of this Letter is to clarify the effects of structural defects, such as vacancies~\cite{rf:hashi} and Stone-Wales (SW) defects~\cite{rf:sw}, on quantized thermal conductance in CNTs. Contact scattering resistance is not considered here to focus on the {\it intrinsic} conductance of defective CNTs.\par

In order to clarify the effects of structural defects, a reliable theory of thermal transport through a nanostructure is needed. Rego and Kirczenow heuristically derived a novel formula for thermal current through a mesoscopic dielectric wire placed between hot and cold heat baths~\cite{rf:rego}:
\begin{eqnarray}
J_{\rm th}=\int_0^\infty\frac{d\omega}{2\pi}\hbar\omega\left[f_L(\omega)-f_R(\omega)\right]\zeta(\omega),
\label{eq:landauer}
\end{eqnarray}
where $f_{L(R)}(\omega)$ is the Bose-Einstein distribution function of equilibrium phonons with energy $\hbar\omega$ in the left (right) lead with temperature $T_L$ ($T_R$), and $\zeta(\omega)$ is the phonon transmission function. The Landauer formalism has some remaining issues to be resolved. First, it is not applicable to an interacting phonon system including phonon-phonon interactions. A second issue is that there is no systematic procedure to obtain $\zeta(\omega)$ for actual nanomaterials with complex structures on an atomic scale. Thus far, $\zeta(\omega)$ has been determined using elastic models for classical sound waves. Of course, such a treatment is not suitable for nanoscale systems with complex atomic structures. In this Letter, we develop a new formalism overcoming these remaining issues by utilizing the nonequilibrium Green's function (NEGF) technique. The advantage of the NEGF formalism is that local physical quantities, such as the nonequilibrium phonon density and the local thermal current, can be calculated. This Letter is a first report of the application of the NEGF formalism to the study of the influence of defect scattering on phonon transport in CNTs.\par

The system Hamiltonian ${\mathscr H}_{\rm sys}$ is described as the sum of the harmonic term 
\begin{eqnarray}
{\mathscr H}_{\rm har}=\!\!\!\!
\sum_{\scriptstyle i\in{\rm sys} \atop \scriptstyle \alpha=xyz}\!\!\!\!
\Big[\frac{p_{i\alpha}^{2}(t)}{2M_i}+\!\!\!\!
\sum_{\scriptstyle j>{i} \atop \scriptstyle \beta=xyz}\!\!\!
\frac{k_{i\alpha, j\beta}}{2}\left(s_{i\alpha}(t)-s_{j \beta}(t)\right)^{2}\Big]
\end{eqnarray} 
and the anharmonic term ${\mathscr H}_{\rm anh}$. Here, $s_{i\alpha}(t)$ is an operator in the Heisenberg picture for atomic displacement from equilibrium along the $\alpha$ direction of the $i$th atom with mass $M_i$. $p_{i\alpha}(t)$ is a momentum operator conjugated to the displacement operator $s_{i\alpha}(t)$, and $k_{i\alpha,j\beta}$ represents the spring constant between the $i$th atom in the $\alpha$ direction and the $j$th atom in the $\beta$ direction. The system Hamiltonian is assumed to be divided into five parts: ${\mathscr H}_{\rm sys}={\mathscr H}_{L}+{\mathscr H}_{LS}+{\mathscr H}_{S}+{\mathscr H}_{RS}+{\mathscr H}_{R}$. Here, ${\mathscr H}_{\rm L/R}$ is the Hamiltonian for the left/right thermal lead, ${\mathscr H}_{\rm S}$ is that for the scattering region, and ${\mathscr H}_{LS(RS)}$ is the Hamiltonian for the coupling between the scattering region and the left (right) lead. The anharmonic term ${\mathscr H}_{\rm anh}$ is also assumed to exist only in the scattering term ${\mathscr H}_{S}$. Different temperatures, $T_L$ and $T_R(<T_L)$, are assigned to the left and right regions of the system, respectively. \par

The thermal current flowing through the interface between the left lead and the scattering region can be calculated from the time evolution of the energy of the left lead:
$J_{\rm th}=-\langle\dot{{\mathscr H}}_{\rm L}\rangle=\frac{{\rm i}}{\hbar}\langle[{\mathscr H}_{\rm L},{\mathscr H}_{\rm sys}]\rangle$, where the bracket $\langle\cdots\rangle$ denotes the nonequilibrium statistical average of physical observable. The thermal current $J_{\rm th}$ is rewritten as 
\begin{eqnarray}
J_{\rm th}=-\lim_{t'\to{t}}\sum_{\scriptstyle i\in{L},j\in{S} \atop \scriptstyle \alpha\beta=xyz}\!\!\!
\frac{k_{i\alpha,j\beta}}{2}\left[{\rm i}\hbar\frac{d}{dt'}G_{i\alpha,j\beta}^{<}(t,t')+
{\rm h.c.}\right]
\end{eqnarray}
using the greater and lesser Green's functions associated with the contact between the left lead and the scattering region: ${\rm i}{\hbar}G_{i\alpha,j\beta}^{>}(t,t')=\left\langle{s}_{i\alpha}(t){s}_{j\beta}(t')\right\rangle$ and ${\rm i}{\hbar}G_{i\alpha,j\beta}^{<}(t,t')=\left\langle{s}_{j\beta}(t'){s}_{i\alpha}(t)\right\rangle$. Since the Green's functions depend only on the time difference in a steady state, it is convenient to work in Fourier space ($\omega$ space). Therefore, in the steady state, the thermal current is expressed as
\begin{eqnarray}
J_{\rm th}
&=&-\int^{\infty}_{0}\! \frac{d\omega}{2\pi}\hbar\omega
\sum_{\scriptstyle i\in{L},j\in{S} \atop \scriptstyle \alpha\beta=xyz}\!\!\!\!\! 
k_{i\alpha,j\beta}\nonumber\\
& &\times\Big[G^{<}_{i\alpha,j\beta}(\omega)-G^{>}_{j\beta,i\alpha}(\omega)+{\rm h.c.}\Big],
\label{eq:I}
\end{eqnarray}
where the relation $G^{<}_{i\alpha,j\beta}(-\omega)=G^{>}_{j\beta,i\alpha}(\omega)$ is used. \par

In accordance with the similar procedure for the NEGF formalism for electronic transport~\cite{rf:haug}, Eq.~(\ref{eq:I}) can be straightforwardly rewritten as 
\begin{eqnarray}
J_{\rm th}=\int^{\infty}_{0}\!\!\frac{d\omega}{2\pi}\hbar\omega
{\rm Tr}\Big[{\bm \Sigma}_L^{>}(\omega){\bm G}^{<}_S(\omega)
-{\bm \Sigma}_L^{<}(\omega){\bm G}^{>}_S(\omega)\Big],
\label{eq:gen}
\end{eqnarray}
where the boldface quantities represent matrices with basis in the scattering region. This is a general expression for the thermal current beyond the Landauer formula (\ref{eq:landauer}) for coherent phonon transport. In Eq.~(\ref{eq:gen}), ${\bm G}^{\llgg}_S(\omega)$ is the greater/lesser Green's function for the scattering region, and ${\bm \Sigma}_{L/R}^{\llgg}(\omega)$ is the greater/lesser self-energy due to coupling to the left/right lead, which is given by
\begin{eqnarray}
{\bm \Sigma}_{L/R}^\llgg(\omega)=-{\rm i}(f_{L/R}(\omega)+1/2\pm{1/2}){\bm \Gamma}_{L/R}(\omega).
\label{eq:lead-sigma}
\end{eqnarray}
Here, ${\bm \Gamma}_{L/R}(\omega)={\rm i}[{\bm\Sigma}_{L/R}^r(\omega)-{\bm\Sigma}_{L/R}^a(\omega)]$, where ${\bm\Sigma}_{L/R}^{r/a}(\omega)$ is the retarded/advanced self-energy due to the coupling to the left/right lead and  is calculated using the mode matching method~\cite{rf:ando} modified for phonon transport. \par

Phonon-phonon scattering is not important for CNTs below 300~K, which is of interest here~\cite{rf:mingo,rf:hone,rf:yu}. In the case of coherent phonon transport, the lesser and greater Green's functions ${\bm G}^{\llgg}$ satisfy the Keldysh equation:
\begin{eqnarray}
{\bm G}^{\llgg}_S(\omega)={\bm G}^{r}_S(\omega)\left(
{\bm \Sigma}_L^{\llgg}(\omega)+{\bm \Sigma}_R^{\llgg}(\omega)
\right){\bm G}^{a}_S(\omega).
\label{eq:keldysh}
\end{eqnarray}
The retarded/advanced Green's function ${\bm G}^{r/a}_S(\omega)$ for the scattering region satisfies the Dyson equation:
\begin{eqnarray}
{\bm G}^{r/a}_S(\omega)=\left[\omega^2{\bm M}-{\bm D}-({\bm \Sigma}_L^{r/a}+{\bm \Sigma}_R^{r/a})\right]^{-1},
\end{eqnarray}
where ${\bm D}$ is the dynamical matrix derived from the second derivative of the total energy with respect to the atom coordinates in the scattering region, and ${\bm M}$ is a diagonal matrix with elements corresponding to the masses of the constituent atoms. In this work, the total energy of the CNTs is determined from the Brenner bond-order potential~\cite{rf:brenner}. \par

Substituting Eqs.~(\ref{eq:lead-sigma}) and (\ref{eq:keldysh}) into Eq.~(\ref{eq:gen}), the thermal current in Eq.~(\ref{eq:gen}) is reduced to the Landauer formula in Eq.~(\ref{eq:landauer}). The phonon transmission function $\zeta(\omega)$ in Eq.~(\ref{eq:landauer}) is also expressed explicitly  as
\begin{eqnarray}
\zeta(\omega)={\rm Tr}\left[{\bm \Gamma}_{L}(\omega){\bm G}^{r}_S(\omega){\bm \Gamma}_{R}(\omega){\bm G}^{a}_S(\omega)\right].
\label{eq:f-l}
\end{eqnarray}
This expression of $\zeta(\omega)$ is equivalent to that derived by Mingo and Yang in a different way~\cite{rf:mingo2}. In the limit of the small temperature difference between the hot and cold heat reservoirs, $T_L-T_R\ll{T}\equiv(T_L+T_R)/2$, the thermal conductance $\kappa\equiv J_{\rm th}/({T_L-T_R})$ is given by 
\begin{eqnarray}
\kappa(T)=\frac{k_B^2T}{h}\int_{0}^{\infty}
dx\frac{x^2{\rm e}^x}{\left({\rm e}^x-1\right)^2}\zeta(k_BTx/\hbar).
\label{eq:kappa}
\end{eqnarray} 
If the phonon transmission is perfect for all acoustic modes in the low-temperature limit $T\to{0}$, Eq.~(\ref{eq:kappa}) is given as a form of an elementary integration that can be integrated analytically, and the thermal conductance is quantized as $M\kappa_0=M(\pi^2k_B^2T/3h)$, where $M$ is the number of acoustic modes. \par

\begin{figure}[t]
  \begin{center}
  \includegraphics[keepaspectratio=true,width=70mm]{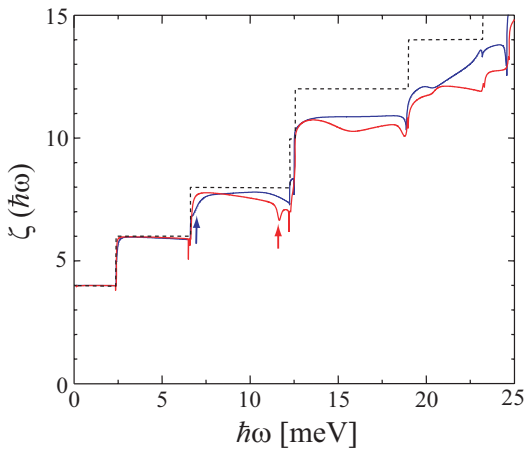}
  \end{center}
  \caption{(color online). Phonon transmission function $\zeta(\hbar\omega)$ for the (8,8)-CNTs. Red and blue curves represent $\zeta(\hbar\omega)$ for the vacancy and SW defect, respectively. The dashed curves are for the perfect CNT. The top-left inset describes the phonon density around the SW defect at 7.0~meV indicated by the blue arrow, and the bottom-right inset is that around the vacancy at 11.6~meV indicated by the red arrow. The red shading on the atom spheres indicates the phonon density. The red shading is not on the same scale for the two insets. (The figures of insets are dropped from this preprint due to large sizes.)}
  \label{fig:1}
\end{figure}

\begin{figure}[t]
  \begin{center}
  \includegraphics[keepaspectratio=true,width=70mm]{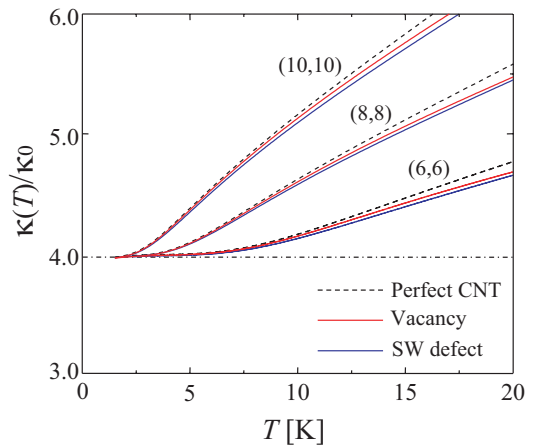}
  \end{center}
  \caption{(color online). Low-temperature thermal conductances in (6,6)-, (8,8)- and (10,10)-CNTs. Red and blue curves represent $\kappa(T)/\kappa_0$ for the CNTs with the vacancy and SW defect, respectively. The dashed curves are for the perfect CNT.}
  \label{fig:2}
\end{figure}

We now apply the formalism developed here to the thermal transport in defective CNTs. Figure~\ref{fig:1} represents the phonon transmission function for the (8,8)-CNT with and without defects. The dashed curve is the transmission function $\zeta_{\rm p}(\hbar\omega)$ for the perfect (8,8)-CNT without any defects, and it displays a clear stepwise structure that gives the number of phonon channels. In the low-energy region below $2.4$~meV, being the energy gap of the lowest optical modes, the dashed curve shows $\zeta_{\rm p}(\hbar\omega)=4$ indicating the number of acoustic branches corresponding to longitudinal, twisting, and doubly degenerated flexural modes. Reflecting the perfect transmission for all acoustic modes, the thermal conductance $\kappa(T)$ shows four universal quanta, $4\kappa_0=4(\pi^2k_B^2T/3h)$, in the low-temperature limit (see Fig.~\ref{fig:2}). \par

The transmission function $\zeta_{\rm vac}(\hbar\omega)$ for the (8,8)-CNT with the vacancy and $\zeta_{\rm SW}(\hbar\omega)$ for the (8,8)-CNT with the SW defect are described by the red and blue curves in Fig.~\ref{fig:1} respectively. The $\zeta_{\rm vac/SW}(\hbar\omega)$ is dramatically deformed from $\zeta_{\rm p}(\hbar\omega)$ owing to defect scattering particularly at high energies. However, it remains unchanged in the low-energy region. This is because the long wavelength acoustic phonons with low energies in the CNTs are not scattered by local defects such as vacancies and SW defects. This leads to the important conclusion that the thermal conductance of CNTs exhibits $4\kappa_0$ at cryostatic temperatures even when the CNTs include a small amount of defects (see Fig.~\ref{fig:2}). \par

The transmission function $\zeta_{\rm vac}(\hbar\omega)$ shows some dips at particular energies, which are clearly distinguished from the dips originating from the van Hove singularity of the optical phonon branches. These dip positions coincide with the peaks in the local density of states (LDOS) around the vacancy (not shown). The appearance of LDOS peaks means that the phonon density is highly localized around the vacancy. The transmission dips arise from the scattering of incident phonons from the lead by the phonon localized states. The bottom-right inset in Fig.~\ref{fig:1} shows the phonon density around the vacancy at 11.6~meV indicated by the red arrow in Fig.~\ref{fig:1}.  The LDOS peak at 11.6~meV is located at the lowest position and largest intensity among the peaks. Similarly, $\zeta_{\rm SW}(\hbar\omega)$ has some dips at particular energies coinciding with positions of the LDOS peak due to the SW defect.
The dip at 7.0~meV indicated by the blue arrow lies at the lowest position among the LDOS peaks associated with the SW defect. As shown in the top-left inset in Fig.~\ref{fig:1}, the phonon density at 7.0~meV is strongly localized around the SW defect. \par

Substituting the $\zeta(\hbar\omega)$ obtained into Eq.~(\ref{eq:kappa}), we can determine the thermal conductance $\kappa(T)$ as a function of temperature $T$. The low-temperature behavior of $\kappa(T)$ normalized to the universal quantum $\kappa_0$ for (6,6)-, (8,8)- and (10,10)-CNTs with and without defects are shown in Fig~\ref{fig:2}. The red (blue) curves represent $\kappa_{\rm vac(SW)}/\kappa_0$ for the CNTs with the vacancy (SW defect). The dashed curves represent $\kappa_{\rm p}/\kappa_0$ for perfect CNTs. As discussed above, all thermal conductance curves approach 4 in the limit of $T\to{0}$, even in the presence of local defects. In the temperature region shown in Fig.~\ref{fig:2}, the thermal conductance $\kappa_{\rm SW}/\kappa_0$ is slightly lower than $\kappa_{\rm vac}/\kappa_0$. This is because propagating phonons in this region are dominantly scattered by the phonon localized state associated with the SW defects at 7.0~meV, and not by the phonon localized states associated with a vacancy at the higher energy of 11.6~meV (see Fig.\ref{fig:1}). \par

\begin{figure}[t]
  \begin{center}
  \includegraphics[keepaspectratio=true,width=70mm]{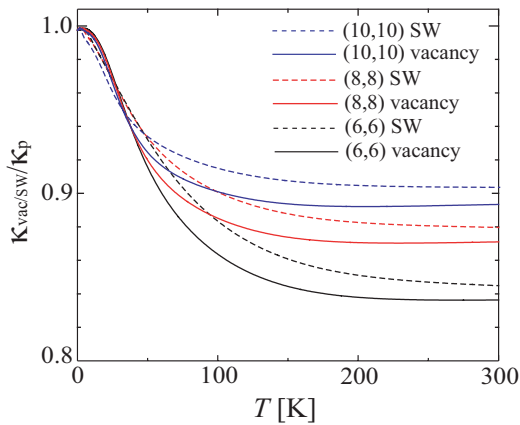}
  \end{center}
  \caption{(color online). The temperature dependence of the ratio $\kappa_{\rm vac(SW)}/\kappa_{\rm p}$ for the CNT with the vacancy (SW defect). The black-, red-, and blue-solid (dashed) curves are $\kappa_{\rm vac(SW)}/\kappa_{\rm p}$ for (6,6)-, (8,8)- and (10,10)-CNTs, respectively. }
  \label{fig:3}
\end{figure}

We next describe the CNT-diameter dependence of defect scattering on thermal conductance for moderate temperatures up to 300~K. Figure~\ref{fig:3} shows the ratios $\kappa_{\rm vac}/\kappa_{\rm p}$ and $\kappa_{\rm SW}/\kappa_{\rm p}$ for (6,6)-, (8,8)- and (10,10)-CNTs as a function of $T$. The black-, red-, and blue-solid (dashed) curves are the ratios $\kappa_{\rm vac(SW)}/\kappa_{\rm p}$ for (6,6)-, (8,8)- and (10,10)-CNTs with the vacancy (SW defect), respectively.  All the $\kappa_{\rm vac(SW)}/\kappa_{\rm p}$ curves decrease rapidly with increasing temperature and become nearly independent of the temperature at $\sim$300~K. The terminal value of $\kappa_{\rm vac(SW)}/\kappa_{\rm p}$ also decreases as the CNT gets thinner. In other words, the influence of defect scattering in {\it thin} CNTs on the thermal conductance is more significant than that in {\it thick} CNTs. Interestingly, $\kappa_{\rm vac}/\kappa_{\rm p}$ is clearly lower than $\kappa_{\rm SW}/\kappa_{\rm p}$ at moderate temperatures, which is in sharp contrast with the low temperature case. That is, the incident phonons are scattered more strongly by the vacancy than by the SW defect. This result is consistent with previous results of classical MD simulations at room temperature~\cite{rf:kondo}. Therefore, the most important finding in this study is that defect scattering, which is responsible for thermal transport, changes remarkably from a quantum to a classical nature with increasing temperature. \par

Finally, we note the electron contribution to the thermal conductance in metallic CNTs. The ballistic electrons in metallic CNTs without any defects contribute four quanta ($=4\kappa_0$) to the thermal conductance~\cite{rf:yama1,rf:mingo}, according to the Wiedemann-Franz relation between electrical conductance and electronic thermal conductance. Even in the presence of the SW defect in metallic CNTs, electronic thermal conductance is quantized as $4\kappa_0$ since the SW defect does not modify the electrical conductance near the Fermi level~\cite{rf:choi,rf:brand}. On the other hand, with the vacancy, the electronic thermal conductance is reduced because the conduction electrons are scattered by the dangling $\sigma$-bond states around the vacancy with an energy close to the Fermi level~\cite{rf:choi}. \par

The influence of defect scattering on {\it phonon} transport in CNTs differs from that on {\it electron} transport. For example, as discussed above, the propagating electrons in CNTs with a vacancy are strongly scattered by the dangling $\sigma$-bond states~\cite{rf:choi}, whereas the long wavelength phonons are not perturbed by the vacancy. \par

In conclusion, we have developed a new formalism for phonon thermal transport in nanostructures using the NEGF technique. Applying this to CNTs with local structural defects, a vacancy and a SW defect, 
we found for the first time that a remarkable change in defect scattering, {\it i.e.}, defect-dependent thermal conductance of CNTs from a quantum to a classical feature, occurs with increasing temperature. Our formalism opens the way for a complete understanding of the underlying physics of phonon transport in nanostructures under various conditions. Finally, we discussed the advantages of our formalism based on the NEGF over other theories.
Our theory, in a straightforward manner based on the Feynmann diagram technique~\cite{rf:valle}, takes into account the phonon-phonon scattering effect that is found to play an important role in thermal transport above room temperature in a recent experiment~\cite{rf:pop}. \par

This work was supported in part by the ``Academic Frontier" Project of MEXT (2005-2010). Part of the numerical calculations was performed on the Hitachi SR11000s at ISSP, The University of Tokyo.

\end{document}